\documentclass[runningheads]{llncs}
\usepackage{graphicx}
\usepackage{amssymb}
\usepackage{xcolor}
\usepackage{cite}

\makeatletter
\usepackage{amssymb}
\newcommand{\ygg@basicalert}[2]{\fbox{\bfseries\sffamily\scriptsize#1}{\sf\small$\blacktriangleright$\color{red}{\textit{#2}}$\blacktriangleleft$}}
\newcommand{\GABRIEL}[1]{\ygg@basicalert{GABRIEL}{#1}}
\newcommand{\CRIS}[1]{\ygg@basicalert{CRISTIANO}{#1}}
\newcommand{\YANN}[1]{\ygg@basicalert{YANN}{#1}}
\newcommand{\FABIO}[1]{\ygg@basicalert{FABIO}{#1}}

\begin{document}

\title{Game Engine Comparative Anatomy}

\author{
Gabriel C. Ullmann\inst{1} \and
Cristiano Politowski\inst{1}\and \\
Yann-Ga\"el Gu\'{e}h\'{e}neuc\inst{1}\and
Fabio Petrillo\inst{2}
}

\authorrunning{G. Ullmann et al.}

\institute{Concordia University, Montreal QC, Canada \\
\email{g\_cavalh@live.concordia.ca, c\_polito@encs.concordia.ca, yann-gael.gueheneuc@concordia.ca}\\
\and
École de Technologie Supérieure, Montreal QC, Canada \\
\email{fabio.petrillo@etsmtl.ca}}

\maketitle

\begin{abstract}
Video game developers use game engines as a tool to manage complex aspects of game development. While engines play a big role in the success of games, to the best of our knowledge, they are often developed in isolation, in a closed-source manner, without architectural discussions, comparison, and collaboration among projects. In this work in progress, we compare the call graphs of two open-source engines: Godot 3.4.4 and Urho3D 1.8. While static analysis tools could provide us with a general picture without precise call graph paths, the use of a profiler such as Callgrind allows us to also view the call order and frequency. These graphs give us insight into the engines' designs. We showed that, by using Callgrind, we can obtain a high-level view of an engine's architecture, which can be used to understand it. In future work, we intend to apply both dynamic and static analysis to other open-source engines to understand architectural patterns and their impact on aspects such as performance and maintenance.

\end{abstract}

\section{Introduction}

Game engines allow developers to create games in an agile and standardized way. While there is a wide array of free, open-source engines, many large studios choose to develop proprietary solutions to be used internally. Famous examples of this approach are Frostbite \footnote{https://www.ea.com/frostbite}, id Tech \footnote{https://arstechnica.com/gaming/2018/08/doom-eternal-reveals-new-powers-puts-hell-back-on-earth/}, and RAGE \footnote{https://www.inverse.com/gaming/gta-6-leaks-new-rage-engine}. There are also popular closed-source engines available publicly, such as Unity and Source \cite{politowski_engines_2021}.

When writing a proprietary engine, developers are able to customize it to fulfill performance or feature requirements of a certain game or game genre, which may give them an edge on competitors. As a downside, however, this development process hinders exchange of information that could be beneficial to the entire game development community. According to \cite{srsen_2021}, ``almost all relevant game engines are closed source which means that particularities on implementation of certain functionalities are not available to developers. The only way for a developer to understand the way certain components work and communicate is to create his/her own computer game engine''.

Developers may also find obstacles when creating a game engine from scratch, especially due to a lack of standards on how high-level architecture components should be created and related for this kind of system. Even though books have been published on the topic of game engine architecture \cite{eberly_3d_2007, gregory_game_2018, lengyel_foundations_2016}, according to \cite{anderson_case_2008}, often these publications ``tend to only briefly describe the high-level architecture before plunging straight down to the lowest level and describing how the individual components of the engine are implemented''. Furthermore, they mention that ``such literature offers an excellent source of information for writing an engine, but provides little assistance for designing one when the requirements are different from the solution described''.

We believe game engines would benefit from experience exchanges with other projects of the same kind. A comparative analysis would allow us to identify commonalities and propose points of improvement to existing engines. In this work in progress, we compare the call graphs of two open-source engines, Godot and Urho3D, considering these aspects.

The Callgrind profiling tool is used to generate these call graphs, which reflect the execution of a base game project produced with each engine. By comparing the call graphs, we aim to observe the main components of each engine and how responsibilities are divided among them. 

While static analysis tools could provide us with a general picture without precise call graph paths, we chose to use a profiler because it allows us to view the call order, frequency, and the number of CPU cycles taken by each method. As a result of this analysis, we will answer the following questions: Are game engine designs similar? If so, how similar are they? Our hypothesis is that game engines follow a similar design and architectural structure.

We show that producing a high-level architecture view of the engine is possible using a profiling tool and that it is a way to get insights into an engine's design. In future work, we will compare these architectures with those proposed by researchers, both through static and dynamic analysis. We will study how design choices made in each case may influence performance, maintenance, and the range of games that can be produced with a given engine.

\section{Related Work}

Several works compare game engines concerning ease of use \cite{dickson_experience-based_2017}, available tools and target platforms, and also to determine which are more suitable for a given platform \cite{pattrasitidecha_comparison_2014} or game genre \cite{pavkov_comparison_2017}. However, these comparisons are made from a game developer point of view and do not encompass details related to engine design or implementation. Other works focus on proposing architecture to fulfill a specific requirement, such as low-energy consumption for mobile devices \cite{christopoulou_overview_2017}. Novel distributed architectures have also been proposed \cite{maggiorini_smash_2016, marin_prototyping_2019}.

In our work, we compare Godot and Urho3D, not simply in terms of what features are available and in what situations they fit best, but also looking at how engine subsystems (e.g., graphics, audio, physics, etc.) are organized, initialized and in which ways they interact and relate.

\section{Approach}
\subsection{Overview}
Our approach consists of five steps. First, we selected C++ engines available on Github (more details on Section \ref{selection}) and cloned their repositories. Following the documentation or running scripts provided by the developer, we compiled them. Using the engine's editor, we created a new project, setting up the minimum necessary to make it run. We will hereby call this our ``base game'' \footnote{https://github.com/gamedev-studies/engine-profiling-projects/tree/master/BaseGame}. A Godot project demands at least one scene object to run, which can be created via the engine's editor. Urho3D, on the other hand, has no editor: one must create a .cpp file and include Urho3D's library to create a new project. We can run the game loop by calling the \textit{Start} method, no scene creation needed. 

We then ran the base game's packaging process, which allows us to obtain an executable that can be analyzed by the profiler. In the case of Urho3D, there is no packaging, only .cpp file compilation. Normally the code would be compiled in release mode, with all optimizations in place, but in our case, we changed the packaging settings to obtain an executable with debug symbols. 

We ran Callgrind on the base game executable to obtain the list of all calls made by the program, which it saves into a log file. Using KCachegrind, we converted this Callgrind log into a visual representation of the program's execution, a call graph. Class and method names in the graph guided our analysis since they helped us understand the responsibilities of each component.
\subsection{Engine selection} \label{selection}
We searched for all repositories on GitHub filtering by topic (\textit{game-engine}) and language (C++). We chose to consider only C++ since this is one of the most popular languages for game engine development \cite{politowski_engines_2021}. After that, we selected 15 projects and ordered them by number of stars and forks. We then proceeded to clone and build each one of them, starting by those with the highest sum of stars and forks. At the time of writing of this paper, we managed to compile and produce a running executable for two of these engines: Godot and Urho3D.

\subsection{Engine compilation and analysis}
We cloned the GitHub repositories for Godot and Urho3D, working with the latest version at the time for each: commit \textit{feb0d} at the \textit{master} branch for Urho3D and commit \textit{242c05} at the \textit{3.4.3-stable} branch for Godot. 

Following the instructions in the documentation, we compiled Godot export templates for Linux (we used Ubuntu 20.04.4 LTS), both in release and debug modes. These are binary files containing engine code compiled for a specific target platform. Before generating an executable for any Godot project, we must link these templates to the project via the editor's GUI. We created a Godot base game, generated the executable, and then ran it with Callgrind for 30 seconds.

\section{Godot's call graph}

\begin{figure}[htb!]
    \centering
    \includegraphics[width=.7\linewidth]{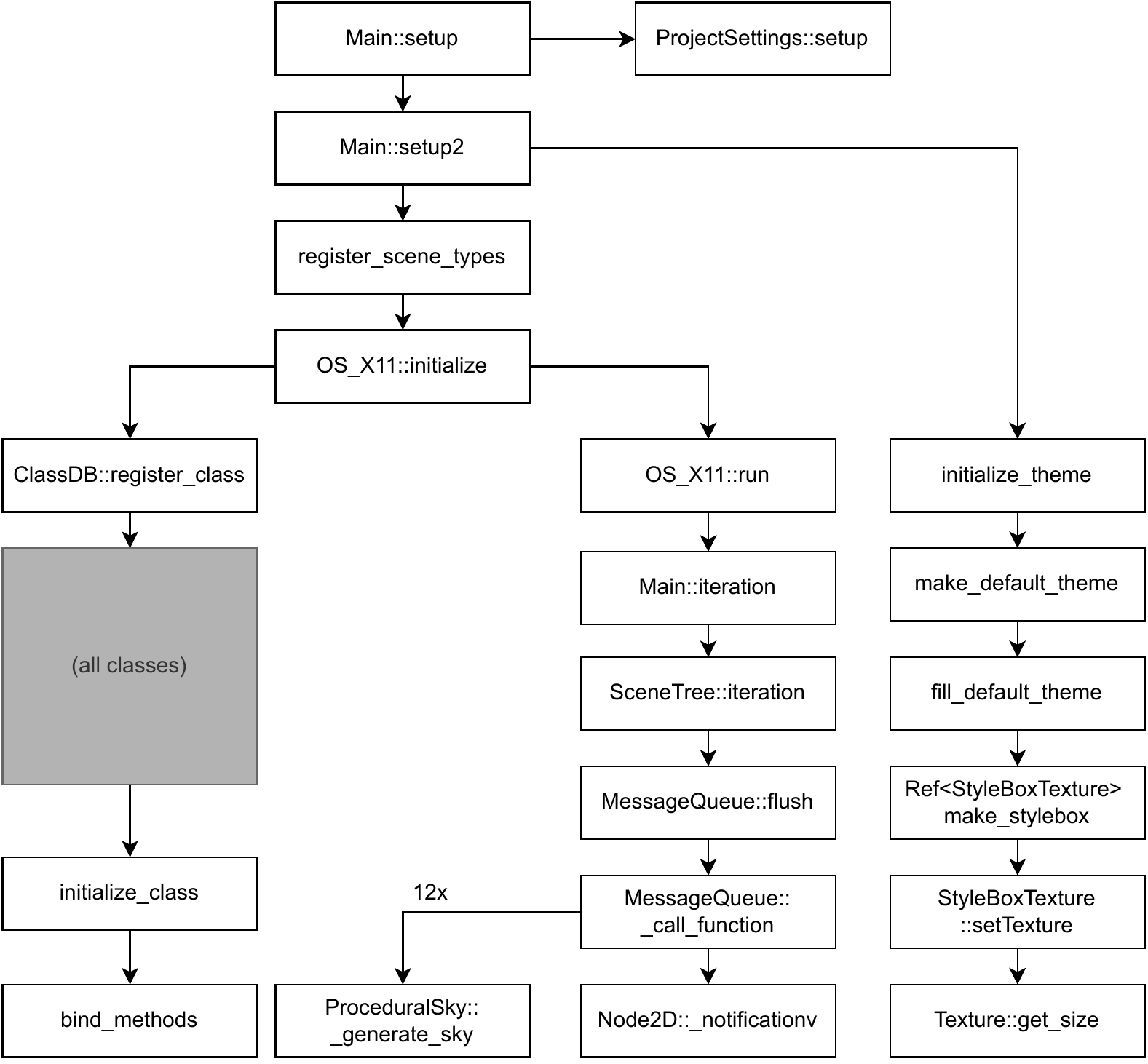}
    \caption{Godot's base game call graph, showing which methods are called.} \label{godot-base}
\end{figure}

We verified that the first tasks Godot executes upon running a game are registering classes and initializing a window (Figure \ref{godot-base}). Since our base game is running on Ubuntu, Godot calls the X11 windowing system. After getting a window instance, the engine then proceeds to call methods to create a scene. Once ready, a notification is sent into the engine's message queue. According to the engine documentation, the \textit{MessageQueue} class is ``a thread-safe buffer to queue set and call methods for other threads'' \footnote{https://docs.godotengine.org/en/stable/development/cpp/custom\_godot\_servers.html}.

Also during startup, Godot's calls methods to set up its GUI theme system \footnote{https://docs.godotengine.org/en/stable/classes/class\_theme.html}, such as \textit{initialize\_theme}, even though in our base game it is not being used. Similarly, a \textit{ProceduralSky::\_generate\_sky} method is called repeatedly even though no sky is drawn. According to Godot's documentation\footnote{https://docs.godotengine.org/en/stable/classes/class\_proceduralsky.html}, this class procedurally generates a sky object, which is ``stored in a texture and then displayed as a background in the scene''. Since no textures are loaded into the base game, nothing is drawn, but the sky generation method is called and updated anyway.

\begin{figure}[htb!]
    \centering
    \includegraphics[width=\linewidth]{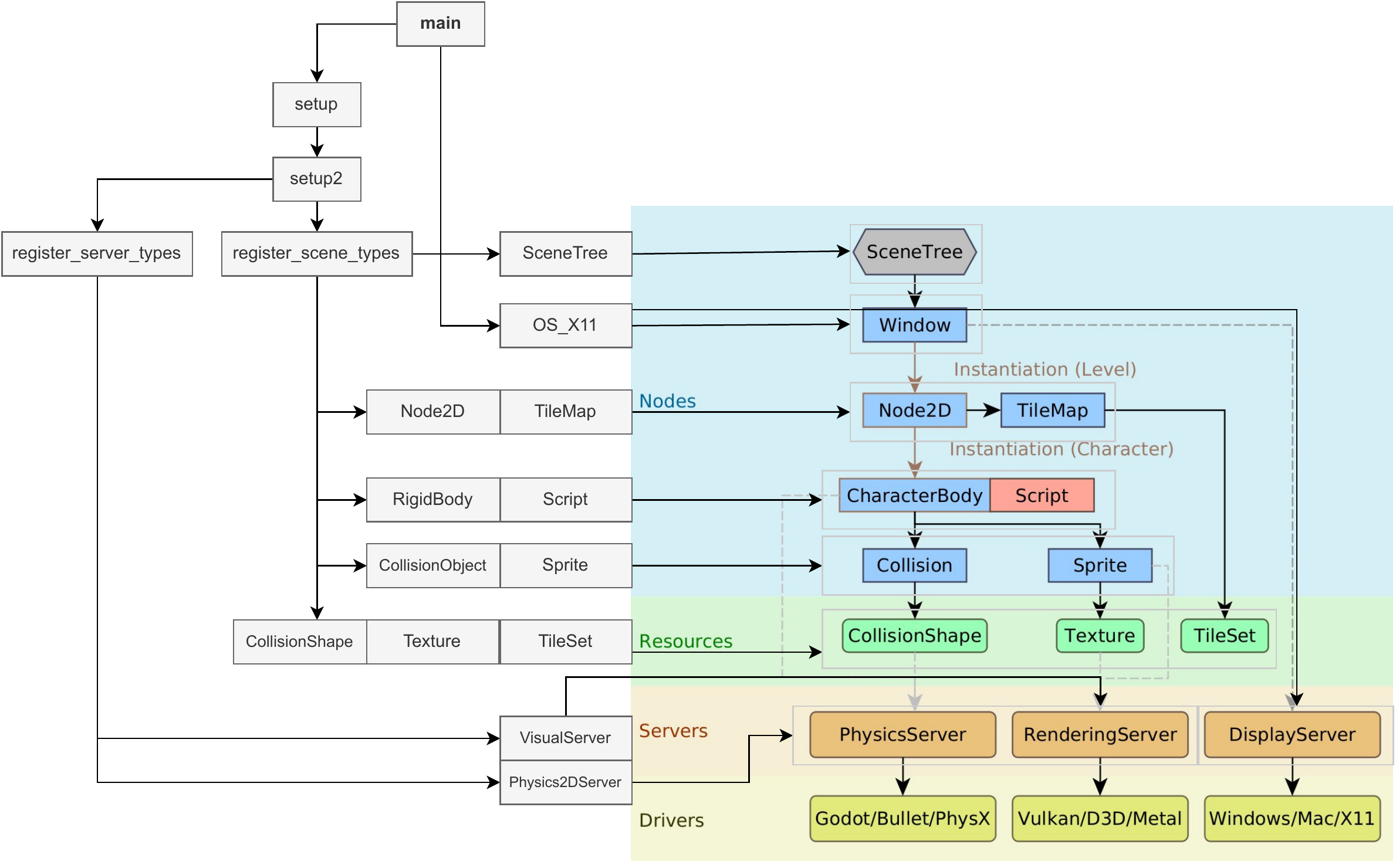}
    \caption{Comparison between the call graph for Godot base game and diagram of ``layers of abstraction'' by Juan Linietsky}
    \label{godot-comparison}
\end{figure}

To validate our approach, we compared Godot's call graph with a ``layers of abstraction'' diagram posted on Twitter by Godot's creator Juan Linietsky \footnote{https://twitter.com/reduzio/status/1506266084420337666}. This comparison can be seen in Figure \ref{godot-comparison}. Searching by class names, we found matches for all classes, even though the names were not the same (e.g \textit{PhysicsServer} and \textit{Physics2DServer}).

The most notorious difference is the class described by Linietsky as \textit{Window}. There is no class with this name on Godot's code base. We believe, however, that he referred to windowing systems in general and therefore decided to simply use the term \textit{Window} since his diagram was platform-agnostic.

Also regarding the windowing system, we found no class named \textit{DisplayServer} on the call graph. To understand what part of our call graph corresponds to this class, we used method names as a reference. Inspecting the source code, we determined that methods \textit{get\_singleton} and \textit{has\_feature} from the \textit{DisplayServer} class were called by the \textit{ProjectSettings} class, which is instantiated by the first \textit{setup} method. In the base game's graph, we found this same method name being called by a class named \textit{OS}. 
\section{Urho3D's call graph}

\begin{figure}[htb!]
    \centering
    \includegraphics[width=\textwidth]{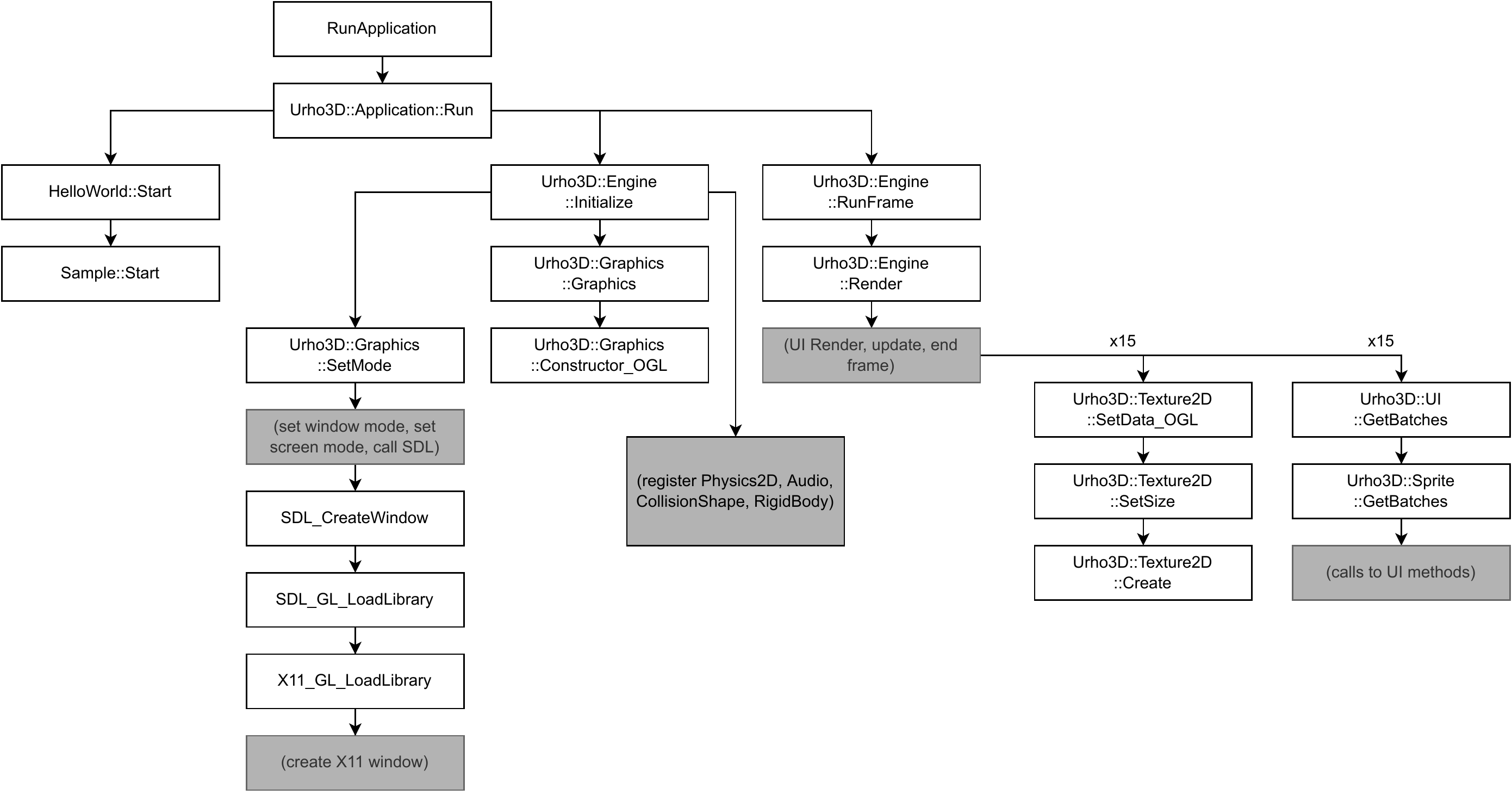}
    \caption{Urho3D's base game call graph, showing which methods are called.} \label{urho3d-base}
\end{figure}

Differently from Godot, Urho3D initializes its graphics object first and only then calls DSL (Simple DirectMedia Layer) and X11 to open a window, as shown in Figure \ref{urho3d-base}. However, they both initialize UI-related code even though there is no UI to draw, and register all game object classes even though they are not instantiated. 

Furthermore, our attention was drawn to the fact that, while Urho3D calls one initialization method per class, Godot calls \textit{Main::setup} and \textit{Main::setup2}. These apparently redundant calls are justified by the developer as a way to separate low and high level ``singletons and core types'' \footnote{Line 342 at https://github.com/godotengine/godot/blob/3.4/main/main.cpp}.

\section{Comparing Godot and Urho3D}

In terms of similarities, both Godot and Urho3D have the same features present in the most widely used game engines, such as graphics, audio, and physics. Both engines register all game object classes upon startup and initialize graphics systems, even though there is nothing more than an empty window to draw.

\begin{figure}[htb!]
    \centering
    \includegraphics[width=\textwidth]{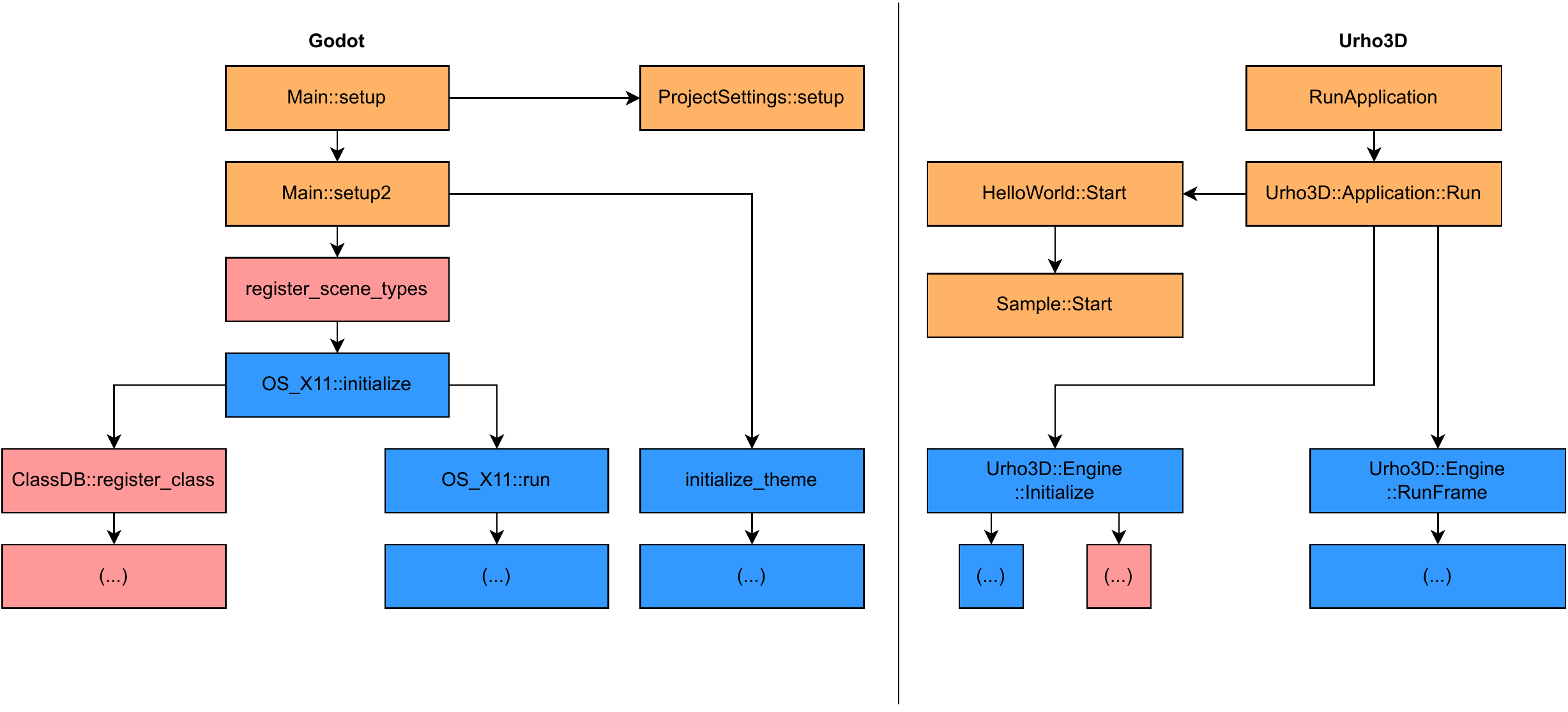}
    \caption{Division of responsibility on Godot and Urho3D: initialization methods (orange), class registration (red) and graphics (blue). Call tree ramifications omitted for brevity. } \label{resp-comp}
\end{figure}

On the other hand, the engines run graphics and window initialization in different orders. Godot creates the window first, Urho3D does the opposite and uses the DSL library to manage the interaction with X11. Godot has a scene manager and internal messaging system which are initialized even in a simple project such as the base game. Urho3D initializes only graphics and UI in this case.

\section{Discussion}
We can use call graphs to visualize the division of responsibilities inside the engines during initialization. They also provide an overview of the most important subsystems and at which moment they are registered and called.

However, these observations do not give us deep insights into architectural patterns and design choices for specific subsystems that could be useful for game engine developers. To obtain this knowledge, we will generate dynamic and static call graphs, as well as C++ \#include graphs for each subsystem in the future.

While comparing the engines, we also identified project and runtime environment conditions that could make the profiler generate the call graph differently. These are important to note since they may change our overall view of the architecture.

\paragraph{Operating System:} There is a possibility that if we ran the profiler on a Windows or Android version of the same base games, we would obtain a different call graph. Each platform has its own system calls, GUI and data types, so the engines account for this by adding platform-specific code. If the call graph would be largely different, this difference could indicate that code evolution and maintainability are compromised.

\paragraph{Active Features:} Important class or method names may not be present in the graphs because they have not been called during the analyzed run. The base game only calls methods from the engine ``core'' and not others referring to more specific features (e.g., networking methods, which would be needed in a multiplayer game).

\paragraph{Other Means of Analysis:} Results could vary when running our base game with other profilers such as \texttt{gprof} or \texttt{gperftools}. Furthermore, while dynamic analysis can provide us with a precise call graph, it only represents a given run of the program and is therefore only a partial view.

\section{Conclusion}

In this work in progress, we showed that producing a high-level architecture view of an engine is possible with the use of a profiler, such as Callgrind, which generates engine call graphs. We compared Godot 3.4.4 and Urho3D 1.8 not only by looking at the engine's subsystems but also in which order and frequency they are called, which would not be possible by applying static analysis alone. 

As for the research question ``are game engine designs similar?'', we concluded that Godot and Urho3D have a similar feature set and therefore could be used to create the same types of games. Their features are the same present in most widely used game engines, but they divide responsibilities and call them in the code in different ways. Also, we identified that both engines initialize subsystems even though they are not used (e.g. UI). While this may consume resources, it seems to be the right choice since these systems are often used.

In future work, we intend to compare different open-source engine call graphs, starting with Unreal Engine 5, to understand what architectural design patterns are most frequently applied to them and how similar they are among engines. Also, we will compare these architectures with those proposed by researchers, using static and dynamic analysis, to understand the impact of design on engine performance, maintenance, and feature richness.

\section*{Acknowledgements}
The authors were partially supported by the NSERC Discovery Grant and Canada Research Chairs programs.

\bibliographystyle{abbrv}
\bibliography{main.bib}

\end{document}